\documentclass[reprint,aps,prd,amsmath,amssymb,superscriptaddress,nofootinbib,floatfix]{revtex4-2}
\usepackage{graphicx}
\usepackage{xcolor}
\usepackage{bm}
\usepackage{amsthm}
\usepackage[colorlinks=true,linkcolor=blue,citecolor=blue,urlcolor=blue]{hyperref}
\usepackage{orcidlink}

\newtheorem{proposition}{Proposition}
\newcommand{\tr}{\mathrm{tr}}
\newcommand{\Pexp}{\mathcal{P}\exp}

\begin{document}

\title{Holonomy classification, gyroscope precession, and Landau levels in G\"odel-type spacetimes and their superfluid analog}

\author{Claudio Furtado\,\orcidlink{0000-0002-3455-4285}}
\email{furtado@fisica.ufpb.br}
\affiliation{Departamento de F\'{\i}sica, CCEN, Universidade Federal da Para\'{\i}ba, Cidade Universit\'{a}ria, 58051-970 Jo\~ao Pessoa, PB, Brazil}

\author{G. Q. Garcia\,\orcidlink{0000-0003-3562-0317}}
\email{gqgarcia99@gmail.com}
\affiliation{Centro de Ci\^encias, Tecnologia e Sa\'ude, Universidade Estadual da Para\'iba, 58233-000 Araruna, PB, Brazil}
\author{Abdullah Guvendi\,\orcidlink{0000-0003-0564-9899}}
\email{abdullah.guvendi@erzurum.edu.tr}
\affiliation{Department of Basic Sciences, Erzurum Technical University, 25050 Erzurum, T\"urkiye}

\author{Alexandre M. de M. Carvalho\,\orcidlink{0009-0006-3540-0364}}
\email{alexandre@fis.ufal.br}
\affiliation{Instituto de F\'{\i}sica, Universidade Federal de Alagoas, Campus A. C. Sim\~oes, Av. Lourival Melo Mota s/n, Tabuleiro do Martins, 57072-970 Macei\'o, AL, Brazil}

\date{\today}

\begin{abstract}
We derive the parallel-transport holonomies of circles centered on the symmetry axis in the homogeneous G\"odel-type family of Rebou\c{c}as and Tiomno, which includes the G\"odel universe, the causal member with $m^2=4\Omega^2$, and the Som--Raychaudhuri limit. The holonomy is governed by a single invariant $\Delta(r)$, whose sign determines whether the Lorentz transformation is elliptic, parabolic, or hyperbolic. We determine the number of parabolic radii across the family and prove that, whenever the azimuthal circles become closed timelike curves (CTCs), the chronology radius satisfies $\Delta(r_c)=-\cosh(mr_c)<0$. We also express this invariant in terms of Killing-vector scalars, derive the vector and spinor Wilson loops, give the exact condition for trivial holonomy after a finite number of turns, and analyze Fermi--Walker transport on circular CTCs. Finally, we discuss an analogy with the effective metric of a rotating superfluid, valid under stated conditions, and derive exact expressions for the radial $L^2$ localization of Klein--Gordon Landau modes relative to the Som--Raychaudhuri chronology radius. The geometric results are exact within the metric family considered; the discussion of the analog system relies on the validity of a coarse-grained effective description.
\end{abstract}

\maketitle

\section{Introduction}

Mandelstam introduced loop variables, the matrices that describe parallel transport along a curve for a given connection, as the basic objects of a path-dependent formulation of gauge theories and gravity~\cite{Mandelstam,Mandelstam1}. Yang~\cite{Yang} and Voronov and Makeenko~\cite{Voronov} developed the idea further. On an open curve a loop variable depends on the curve and on the frames chosen at its end points. On a closed curve it defines the holonomy, whose conjugacy class does not depend on the frame. The trace of the holonomy, the gravitational Wilson loop~\cite{Wilson}, therefore carries frame-independent information about the global structure of spacetime. Holonomies are also the basic variables of loop quantum gravity~\cite{Ashtekar,RovelliSmolin1,RovelliSmolin2}.

Gravitational holonomies are known for Kerr~\cite{BGT}, Schwarzschild--Droste~\cite{Rothman}, Taub--NUT~\cite{Bini}, rotating black strings~\cite{CarvalhoBS}, black cigars~\cite{CarvalhoCigar}, Friedmann--Robertson--Walker cosmologies~\cite{CarvalhoFRW}, and spacetimes with topological defects~\cite{Burgers,Bezerra,FordVilenkin,Assis1,Assis2,BezerraLetelier,FurtadoKK}. In the Schwarzschild case Rothman, Ellis, and Murugan~\cite{Rothman} found what they called band holonomy invariance: on certain circular orbits the holonomy returns to the identity after a finite number of turns.

G\"odel-type spacetimes~\cite{Godel} are an obvious place to look next. They are homogeneous and rotating, and they contain closed timelike curves (CTCs)~\cite{Hawking}. Rebou\c{c}as and Tiomno~\cite{RT} showed that the homogeneous G\"odel-type metrics form a family labeled by the ratio of the vorticity $\Omega$ to a curvature scale $m$. The family contains the original G\"odel universe, a causal solution that is locally $AdS_3\times\mathbb{R}$, and, in the limit $m\to0$, the Som--Raychaudhuri (SR) metric~\cite{SomRay}. These geometries have reappeared in string theory~\cite{HorowitzTseytlin,RussoTseytlin1,RussoTseytlin2,Boyda,Harmark}, and their quantum mechanics is closely tied to the Landau problem~\cite{DFS1,DFS2,DasGegenberg,FST,CarvalhoEPJC,GarciaTorsion,GarciaWeyl,VitoriaSR,BouzenadaSR}. They also appear as analog geometries. Chapline and Mazur~\cite{ChaplineMazur} argued that a superfluid with a dense vortex lattice in rigid rotation is described on large scales by a metric of SR type.

Transport in G\"odel-type spacetimes has been studied before. Loop variables and Wilson loops in the G\"odel universe were studied in Ref.~\cite{Edinelson}, gyroscope precession along geodesics was worked out in Ref.~\cite{BiniGodel}, and Nolan proved a general theorem on when gyroscopes carried along CTCs return to their initial orientation, working out circular CTCs in G\"odel, Som--Raychaudhuri, and several other spacetimes~\cite{Nolan}. As far as we know, nobody has classified the circular holonomies of the whole Rebou\c{c}as--Tiomno family in closed form, or asked how that classification relates to the causal structure.

Our main result is a closed-form classification of the holonomies of centered circles throughout the homogeneous G\"odel-type family. The Lorentz holonomy is controlled by a single invariant, $\Delta(r)$, which determines whether transport is elliptic, parabolic, or hyperbolic. We identify the parameter ranges with different numbers of parabolic radii and show that the chronology radius, whenever it exists, lies in a hyperbolic band and satisfies $\Delta(r_c)=-\cosh(mr_c)$ (Propositions~\ref{prop1} and~\ref{prop2}). We relate this result to the Killing vectors that generate the circular transport, and we also obtain the vector and spinor Wilson loops, the conditions for trivial holonomy after a finite number of turns, and the Fermi--Walker transport of gyroscopes on circular CTCs. We then separate the geometric statements, which hold exactly, from their conditional interpretation in a rotating-superfluid analog, and we quantify the radial $L^2$ localization of scalar Landau modes relative to the Som--Raychaudhuri chronology radius.

We use the signature $(+,-,-,-)$, tetrad indices $a,b,\ldots\in\{0,1,2,3\}$, and units with $c=G=\hbar=1$. In the analog model we keep the sound speed $c_s$ explicit. The paper is organized as follows. Section~\ref{sec:geometry} introduces the G\"odel-type metrics, their causal structure, and the tetrad used throughout, and Sec.~\ref{sec:loops} defines the loop variables. Section~\ref{sec:circles} develops the closed-form classification of the holonomy, establishes its relation to the chronology horizon, and derives the Wilson loops and the conditions for trivial holonomy; Sec.~\ref{sec:open} treats transport along open paths. Section~\ref{sec:FW} analyzes gyroscope precession on circular CTCs. Section~\ref{sec:analogue} discusses the scope and limitations of the superfluid analogy, Sec.~\ref{sec:KG} presents the localization of scalar Landau modes, and Sec.~\ref{sec:conclusions} summarizes our conclusions.

\section{G\"odel-type spacetimes}\label{sec:geometry}

In cylindrical coordinates $(t,r,\phi,z)$ the homogeneous G\"odel-type metrics read~\cite{RT}
\begin{equation}\label{metric}
ds^2=\big[dt+H(r)\,d\phi\big]^2-D^2(r)\,d\phi^2-dr^2-dz^2,
\end{equation}
with
\begin{equation}\label{HD}
H(r)=\frac{4\Omega}{m^2}\sinh^2\!\frac{mr}{2},\qquad D(r)=\frac{\sinh mr}{m},
\end{equation}
so that $H'(r)=2\Omega D(r)$. Here $\Omega$ is the vorticity of the congruence $\partial_t$ and $m$ is a curvature scale. We will mostly use the dimensionless combinations
\begin{equation}\label{kC}
k\equiv\frac{2\Omega^2}{m^2},\qquad C\equiv\cosh mr\ \ (\ge1).
\end{equation}
The G\"odel universe has $k=1$ ($m^2=2\Omega^2$) and the causal Rebou\c{c}as--Tiomno solution has $k=1/2$ ($m^2=4\Omega^2$). The Som--Raychaudhuri spacetime is the limit $k\to\infty$, that is, $m\to0$ at fixed $\Omega$:
\begin{equation}\label{SR}
ds^2=\big(dt+\Omega r^2d\phi\big)^2-r^2d\phi^2-dr^2-dz^2.
\end{equation}

\emph{Causal structure.} Since $g_{\phi\phi}=H^2-D^2$, the circles of constant $t$, $r$, and $z$ become timelike, and hence CTCs, when $\tanh^2(mr/2)>m^2/4\Omega^2$. In terms of $k$ and $C$ the condition reads
\begin{equation}\label{Cc}
C>C_c\equiv\frac{2k+1}{2k-1}\qquad (k>1/2),
\end{equation}
and there are no CTCs when $k\le1/2$. We call the circle $r=r_c$ the chronology horizon. For G\"odel, $C_c=3$, or equivalently $\sinh(mr_c/2)=1$. For SR, $\Omega r_c=1$.

\emph{Tetrad and connection.} We work with the corotating tetrad
\begin{equation}\label{tetrad}
e^0=dt+H\,d\phi,\quad e^1=dr,\quad e^2=D\,d\phi,\quad e^3=dz,
\end{equation}
for which $de^0=2\Omega\,e^1\wedge e^2$ and $de^2=(D'/D)\,e^1\wedge e^2$. The first Cartan equation $de^a+\omega^a{}_b\wedge e^b=0$ gives the non-vanishing connection 1-forms
\begin{subequations}\label{omega}
\begin{align}
\omega^0{}_1&=\omega^1{}_0=\Omega\,e^2=\Omega D\,d\phi,\\
\omega^0{}_2&=\omega^2{}_0=-\Omega\,e^1=-\Omega\,dr,\\
\omega^1{}_2&=-\omega^2{}_1=\Omega\,e^0-D'\,d\phi=\Omega\,dt+(\Omega H-D')\,d\phi.
\end{align}
\end{subequations}
We write $\omega^a{}_b=(\Gamma_\mu)^a{}_b\,dx^\mu$. Since $e^3$ decouples, it is enough to keep the $(0,1,2)$ block:
\begin{subequations}\label{Gammas}
\begin{align}
\Gamma_\phi&=\begin{pmatrix}0&A&0\\ A&0&B\\ 0&-B&0\end{pmatrix},\\
\Gamma_t&=\begin{pmatrix}0&0&0\\ 0&0&\Omega\\ 0&-\Omega&0\end{pmatrix},\quad
\Gamma_r=\begin{pmatrix}0&0&-\Omega\\ 0&0&0\\ -\Omega&0&0\end{pmatrix},
\end{align}
\end{subequations}
with
\begin{equation}\label{AB}
A=\Omega D=\frac{\Omega}{m}\sinh mr,\quad B=\Omega H-D'=(k-1)C-k .
\end{equation}
All three matrices lie in the Lie algebra $so(2,1)$ spanned by the boosts $K_1$, $K_2$ and the rotation $J_3$ acting on $(e^0,e^1,e^2)$. In the SR limit, $A=\Omega r$ and $B=\Omega^2r^2-1$. We checked Eqs.~(\ref{omega})--(\ref{AB}), and every result below, with computer algebra; the scripts are in the Supplemental Material~\cite{SM}.

\section{Loop variables}\label{sec:loops}

Along a curve $\gamma$ running from $x(\lambda_1)$ to $x(\lambda_2)$, the tetrad components of a parallel-transported vector obey $dV^a/d\lambda=-(\Gamma_\mu)^a{}_b\,\dot x^\mu V^b$. The loop variable is the path-ordered exponential
\begin{equation}\label{U}
U(\gamma)=\Pexp\!\left(-\int_\gamma\Gamma_\mu\,dx^\mu\right),
\end{equation}
Under a local Lorentz transformation $\Lambda(x)$ of the tetrad it changes as $U\to\Lambda(x_2)\,U\,\Lambda^{-1}(x_1)$. On a closed curve $U$ is the holonomy, and its conjugacy class, and with it the Wilson loop $W(\gamma)=\tr\,U(\gamma)$, is gauge invariant. On an open curve (Sec.~\ref{sec:open}) $U$ also reflects the choice of frame, so it has to be read with care.

\section{Holonomy of centered circles}\label{sec:circles}

\subsection{Closed form and $SO(2,1)$ classification}

Take the circle $\gamma_r$ of constant $t$, $r$, and $z$, with $0\le\phi\le2\pi$. Because $\Gamma_\phi$ does not depend on $\phi$, the path ordering is trivial and $U(\gamma_r)=\exp(-2\pi\Gamma_\phi)$. Multiplying out the matrices gives
\begin{equation}\label{cube}
\Gamma_\phi^3=-\Delta\,\Gamma_\phi,\qquad \Delta(r)\equiv B^2-A^2=-\tfrac12\tr\,\Gamma_\phi^2 .
\end{equation}
Splitting the exponential series into odd and even powers then gives, for $n$ turns,
\begin{equation}\label{Uclosed}
U^n(\gamma_r)=\openone-\frac{\sin(2\pi n\sqrt{\Delta})}{\sqrt{\Delta}}\,\Gamma_\phi+\frac{1-\cos(2\pi n\sqrt{\Delta})}{\Delta}\,\Gamma_\phi^2 ,
\end{equation}
for $\Delta>0$. When $\Delta<0$ we replace $\sqrt{\Delta}$ by $i\sqrt{-\Delta}$, which turns $\sin$ and $\cos$ into $\sinh$ and $\cosh$. When $\Delta=0$ the series stops at second order, $U^n=\openone-2\pi n\,\Gamma_\phi+2\pi^2n^2\,\Gamma_\phi^2$.

The vector
\begin{equation}\label{axis}
n^a=(B,\,0,\,-A,\,0),\qquad \eta_{ab}n^an^b=\Delta,
\end{equation}
spans the kernel of $\Gamma_\phi$, so $U(\gamma_r)$ leaves it unchanged. Its norm sorts the holonomy into three types:
\begin{itemize}
\item \emph{elliptic} ($\Delta>0$): a rotation by $2\pi\sqrt{\Delta}$ about the \emph{timelike} axis $n^a$. An observer with four-velocity $n^a/\sqrt{\Delta}$ sees a pure rotation.
\item \emph{parabolic} ($\Delta=0$): a null rotation about the null vector $n^a$.
\item \emph{hyperbolic} ($\Delta<0$): a boost of rapidity $2\pi\sqrt{-\Delta}$ in the plane orthogonal to the \emph{spacelike} axis $n^a$.
\end{itemize}
In the tetrad basis the matrix (\ref{Uclosed}) mixes $e^0$, $e^1$, and $e^2$. It is a plain rotation in the $(e^1,e^2)$ plane only where $A=0$, which happens on the axis.

\emph{Axis limit and relative rotation angle.} As $r\to0$, $A\to0$ and $B\to-1$, so $\Delta\to1$ and the vector holonomy tends to the identity. The $2\pi$ rotation contained in $\Gamma_\phi$ in this limit only reflects the turning of the polar tetrad, and it must be distinguished from a physical rotation relative to the frame on the axis. On elliptic orbits a convenient measure of the latter is the relative angle
\begin{equation}\label{deficit}
\delta(r)=2\pi\big(1-\sqrt{\Delta(r)}\big)\pmod{2\pi}.
\end{equation}
This quantity plays the role of the deficit angle of conical geometries~\cite{Bezerra,FordVilenkin}. For a general orbit it is an effective relative angle and does not imply that the geometry is globally conical.

\subsection{The invariant $\Delta$ across the family}

From Eq.~(\ref{AB}) and $\Omega^2\sinh^2mr/m^2=k(C^2-1)/2$ we get
\begin{equation}\label{Delta}
\Delta=\big[(k-1)C-k\big]^2-\frac{k}{2}\,(C^2-1),
\end{equation}
or, as a quadratic in $C$,
\begin{equation}\label{DeltaC}
\Delta=\tfrac12(k-2)(2k-1)\,C^2-2k(k-1)\,C+k^2+\tfrac{k}{2}.
\end{equation}
A quarter of its discriminant is $k(5k-2)/4$, and $\Delta=1$ at $C=1$ for every $k$. Three members of the family give especially simple results:
\begin{align}
k=\tfrac12:&\quad \Delta=\tfrac12(1+C)=\cosh^2\!\tfrac{mr}{2},\label{Dads}\\
k=1\ (\text{G\"odel}):&\quad \Delta=\tfrac12(3-C^2),\label{Dgodel}\\
k\to\infty\ (\text{SR}):&\quad \Delta=(x^2-x-1)(x^2+x-1),\label{DSR}
\end{align}
with $x\equiv\Omega r$ in the SR case.

\begin{proposition}\label{prop1}
(a) For $0<k\le1/2$ the circular holonomy is elliptic at every radius. (b) For $1/2<k\le2$ there is exactly one parabolic radius $r_p$, and the holonomy is hyperbolic for all $r>r_p$. (c) For $k>2$ there are two parabolic radii $r_-<r_+$: the holonomy is elliptic for $r<r_-$, hyperbolic for $r_-<r<r_+$ and elliptic again for $r>r_+$.
\end{proposition}

\begin{proof}
Equation~(\ref{DeltaC}) is a quadratic in $C$ with leading coefficient $a_k=\frac12(k-2)(2k-1)$, vertex $C_*=2k(k-1)/[(k-2)(2k-1)]$, discriminant $k(5k-2)$ and $\Delta(1)=1$. (a) For $k=1/2$, Eq.~(\ref{Dads}) is positive. For $2/5<k<1/2$, $a_k>0$ and $C_*<0$, so both real roots lie below $C=1$. For $0<k<2/5$ the discriminant is negative. At $k=2/5$ it vanishes, and the double root $C_*=-3/2$ lies outside the physical domain $C\geq1$. Hence $\Delta>0$ at every radius when $0<k\leq1/2$. (b) For $1/2<k<2$, $a_k<0$ and $\Delta(1)>0$, so exactly one root has $C>1$. For $k=2$, $\Delta=5-4C$ vanishes only at $C=5/4$. (c) For $k>2$, $a_k>0$, the roots are real, and $C_*>1$ (which is equivalent to $3k>2$). Since $\Delta(1)>0$, both roots then lie in $C>1$.
\end{proof}

For G\"odel the single parabolic radius is at $\cosh mr_p=\sqrt3$. For SR the two parabolic radii are given by the golden ratio $\varphi=(1+\sqrt5)/2$:
\begin{equation}\label{golden}
\Omega r_-=\varphi^{-1}\simeq0.618,\qquad \Omega r_+=\varphi\simeq1.618 .
\end{equation}

\begin{proposition}\label{prop2}
For every $k>1/2$, at the radius where CTCs appear,
\begin{equation}\label{DeltaRc}
\Delta(r_c)=-C_c=-\cosh mr_c<0 .
\end{equation}
The circle where CTCs first appear therefore always lies inside a hyperbolic band.
\end{proposition}

\begin{proof}
At $C=C_c=(2k+1)/(2k-1)$ one has $B=-1/(2k-1)$ and $A^2=4k^2/(2k-1)^2$, so $\Delta=(1-4k^2)/(2k-1)^2=-(2k+1)/(2k-1)$. A coordinate-free proof is given in Sec.~\ref{sec:killing}.
\end{proof}

\begin{figure}[t]
\centering
\includegraphics[width=\columnwidth]{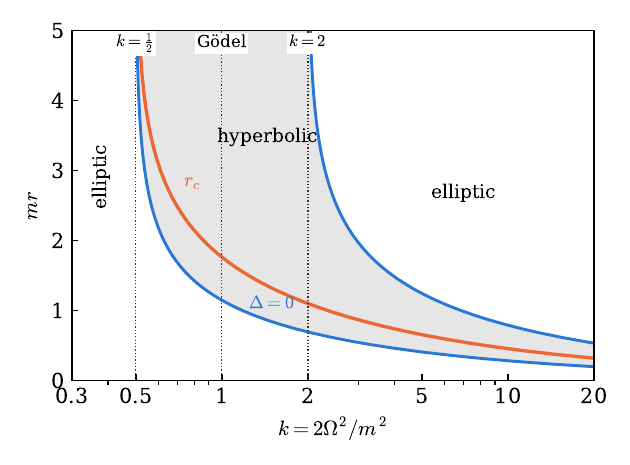}
\caption{Classification of the circular holonomy in the Rebou\c{c}as--Tiomno family. Shaded: hyperbolic ($\Delta<0$); white: elliptic ($\Delta>0$); blue lines: parabolic radii ($\Delta=0$). The orange line is the CTC radius $r_c(k)$, which always lies inside the hyperbolic band (Proposition~\ref{prop2}). Dotted lines mark the causal solution ($k=1/2$), G\"odel ($k=1$), and the transition at $k=2$. SR is the limit $k\to\infty$.}
\label{fig:phase}
\end{figure}

In the SR limit Eq.~(\ref{DeltaRc}) gives $\Delta(r_c)=-1$, in agreement with $\varphi^{-1}<\Omega r_c=1<\varphi$. Equation~(\ref{axis}) makes the result easy to picture. At $r_c$ the circle itself is null, while the axis of its holonomy is spacelike. A vector carried once around the edge of the causal region is therefore boosted, not rotated. Figure~\ref{fig:phase} shows the full picture in the $(k,mr)$ plane.

\subsection{Why the chronology horizon is hyperbolic}\label{sec:killing}

The computation above hides a simple geometric fact. Because the tetrad (\ref{tetrad}) does not depend on $\phi$, a vector with constant tetrad components is Lie-dragged along $\xi=\partial_\phi$, so $\nabla_\xi V=\nabla_V\xi$. Comparing with $\nabla_\xi V^a=(\Gamma_\phi)^a{}_bV^b$ gives
\begin{equation}\label{bivector}
(\Gamma_\phi)^a{}_b=\nabla_b\,\xi^a ,
\end{equation}
so the circular holonomy is $U(\gamma_r)=\exp(-2\pi\nabla\xi)$, the exponential of the Killing bivector of the rotation. The invariant of Eq.~(\ref{cube}) is then $\Delta=\frac12\nabla_a\xi_b\nabla^a\xi^b$. It can be written in terms of the two Killing vectors of the problem, $\xi=\partial_\phi$ and the unit timelike $u=\partial_t$. Using $H=k(C-1)/\Omega$, $D'=C$, and $D'^2-m^2D^2=1$, one finds
\begin{equation}\label{DeltaKilling}
\Delta=1-2\Omega\,(u\cdot\xi)+(\Omega^2-m^2)\,(\xi\cdot\xi),
\end{equation}
with $u\cdot\xi=H$ and $\xi\cdot\xi=H^2-D^2$. Both scalars are invariant, so Eq.~(\ref{DeltaKilling}) holds in any frame.

Proposition~\ref{prop2} now follows at once. At the chronology horizon $\xi$ is null, the last term drops out, and $\Delta(r_c)=1-2\Omega H(r_c)$. On a null circle $H=D$, and for the whole family $\Omega D(r_c)=2k/(2k-1)$, which gives $\Delta(r_c)=-(2k+1)/(2k-1)$. The term $u\cdot\xi=H$ also has a direct physical meaning: $2\pi H$ is the time gap that a clock synchronization around the circle fails to close, the same quantity that produces the Sagnac phase of Sec.~\ref{sec:analogue}. The holonomy at the horizon is a boost because $2\Omega H(r_c)=4k/(2k-1)$ is always larger than 2, far above the value 1 at which $\Delta$ changes sign: by the time a circle becomes null, the synchronization gap is too large for parallel transport to act as a rotation. In the SR limit, Eq.~(\ref{DeltaKilling}) reduces to $\Delta=1-2\Omega^2r^2+\Omega^2(\Omega^2r^4-r^2)$, which reproduces Eq.~(\ref{DSR}).

\subsection{Band holonomy invariance}

On an elliptic orbit $U^n(\gamma_r)=\openone$ if and only if
\begin{equation}\label{band}
n\sqrt{\Delta(r)}=p\in\mathbb{Z},
\end{equation}
This is the band holonomy invariance of Ref.~\cite{Rothman} (see also Ref.~\cite{CarvalhoFRW} for FRW) in exact form. Hyperbolic bands contain no such orbits. For SR, with $u=x^2$ and $q=p/n$, Eq.~(\ref{band}) becomes $u^2-3u+1-q^2=0$, whose solutions are
\begin{equation}\label{bandSR}
\Omega r_{p/n}=\sqrt{\tfrac12\Big(3\pm\sqrt{5+4q^2}\Big)},
\end{equation}
The minus sign (inner elliptic band) needs $q<1$; the plus sign (outer band) works for any $q>0$. The first nontrivial orbit that closes after one turn ($p=n=1$) is at $\Omega r=\sqrt3$. In G\"odel only the inner band exists, and the condition becomes $\cosh^2mr=3-2q^2$ with $0<q<1$.

\subsection{Wilson loops and spinor holonomy}

Taking the trace of Eq.~(\ref{Uclosed}) with $\tr\,\Gamma_\phi^2=-2\Delta$ gives the vector Wilson loop
\begin{equation}\label{W1}
W_1(\gamma_r)=2+2\cos\big(2\pi\sqrt{\Delta}\big)=4\cos^2\!\big(\pi\sqrt{\Delta}\big),
\end{equation}
which becomes $4\cosh^2(\pi\sqrt{-\Delta})$ on hyperbolic orbits.

\begin{figure*}[t]
\centering
\includegraphics[width=0.85\textwidth]{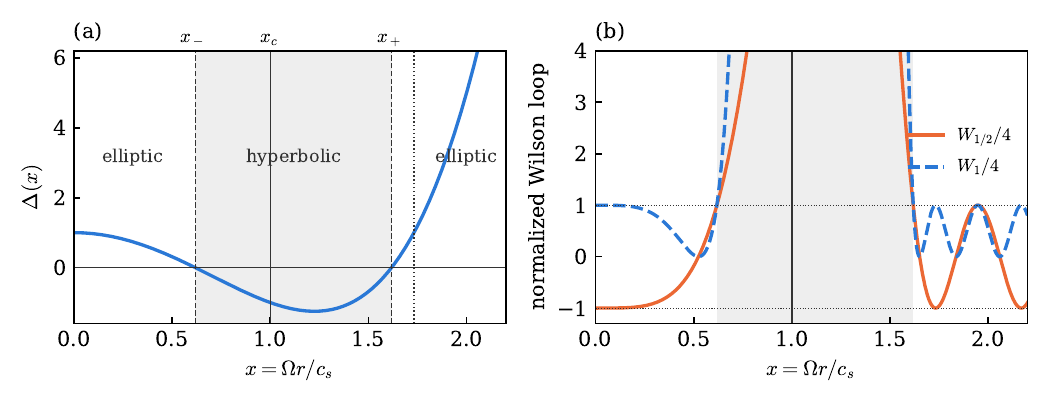}
\caption{Som--Raychaudhuri geometry, or its acoustic analog with $x=\Omega r/c_s$. (a) The holonomy invariant $\Delta(x)$. The hyperbolic band (shaded) runs between the golden-ratio radii $x_\mp=\varphi^{\mp1}$ and contains the CTC radius $x_c=1$. The dotted line marks the first band-invariant orbit, $x=\sqrt3$. (b) Normalized spinor ($W_{1/2}/4$, solid) and vector ($W_1/4$, dashed) Wilson loops.}
\label{fig:SR}
\end{figure*}

For spinors the connection is $\Gamma^{(1/2)}_\mu=\frac18\,\omega_{ab\mu}[\gamma^a,\gamma^b]$. It satisfies $(\Gamma^{(1/2)}_\phi)^2=-\frac{\Delta}{4}\,\openone$, so
\begin{equation}\label{Uspin}
U_{1/2}(\gamma_r)=\cos\big(\pi\sqrt{\Delta}\big)\,\openone-\frac{2\sin(\pi\sqrt{\Delta})}{\sqrt{\Delta}}\,\Gamma^{(1/2)}_\phi ,
\end{equation}
and
\begin{equation}\label{W12}
W_{1/2}(\gamma_r)=4\cos\big(\pi\sqrt{\Delta}\big).
\end{equation}
As a check, $W_1=|W_{1/2}/2|^2$, as it must be for the $(\frac12,\frac12)$ representation. Spinors differ from vectors in two ways. First, $U_{1/2}\to-\openone$ as $r\to0$: this is the usual sign picked up by a spinor when the polar frame turns by $2\pi$, and the physical quantity is the ratio $U_{1/2}(r)\,U_{1/2}^{-1}(0)$, which depends only on the deficit (\ref{deficit}). Second, a vector comes back to itself when $\sqrt{\Delta}\in\mathbb{Z}$, but a spinor needs $\sqrt{\Delta}\in2\mathbb{Z}$. At the SR orbit $\Omega r=\sqrt3$, for example, the vector holonomy is trivial while $U_{1/2}=-\openone$. The band-invariant orbits for spinors are therefore a subset of those in Eq.~(\ref{band}). Figure~\ref{fig:SR} shows $\Delta$ and both Wilson loops for SR.

\section{Open paths: time and radial transport}\label{sec:open}

\emph{Time displacement.} Along a line of constant $r$, $\phi$, and $z$, $\Gamma_t$ is constant and $U(\gamma_t)=\exp(-\tau\Gamma_t)$ is a rotation by $\Omega\tau$ in the $(e^1,e^2)$ plane. These lines are the worldlines of the comoving observers, and they are geodesics because $\partial_t$ is a Killing vector with $g_{tt}=1$. Parallel transport and Fermi--Walker transport therefore agree, and the result says that a gyroscope carried by a comoving observer precesses at angular velocity $\Omega$ relative to the corotating triad. This is the vorticity of the congruence, G\"odel's ``rotation of matter relative to the compass of inertia''~\cite{Godel}. It takes the same value at every radius, on either side of $r_c$, so it comes from the vorticity and has nothing to do with the CTCs.

\emph{Radial displacement.} Along a line of constant $t$, $\phi$, and $z$, $\Gamma_r$ is constant and $U(\gamma_{\rm rad})=\exp(\Omega\,\Delta r\,K_2)$ is a boost of rapidity $\Omega\Delta r$ in the $(e^0,e^2)$ plane. Because the path is open, this boost depends on our choice of the corotating tetrad (\ref{tetrad}). What it does show is that a tetrad adapted to $e^0$ cannot be parallel along radial lines, which follows directly from $de^0\neq0$.

\section{Gyroscopes on circular CTCs}\label{sec:FW}

\begin{figure}[t]
\centering
\includegraphics[width=0.95\columnwidth]{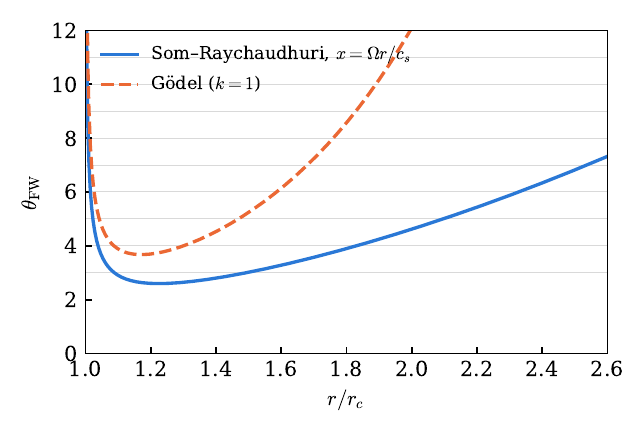}
\caption{Fermi--Walker rotation number $\theta$ per loop of a gyroscope on a circular CTC, Eqs.~(\ref{thetaSR})--(\ref{thetaG}), as a function of $r/r_c$. Gray lines mark integer values, where the gyroscope is periodic. The two curves agree with Figs.~1 and 2 of Ref.~\cite{Nolan}.}
\label{fig:FW}
\end{figure}

For $r>r_c$ the circle $\gamma_r$ is timelike, so an accelerated observer can follow it. A gyroscope carried by this observer is Fermi--Walker (FW) transported rather than parallel transported. FW holonomies along circular orbits were studied before in cosmic-string and Schwarzschild--AdS spacetimes, where the acceleration produces an analog of Thomas precession~\cite{BakkeFW}. For a curve with tangent $T$, $\epsilon=\mathrm{sign}(T\cdot T)$, unit tangent $u$, and acceleration $a=\nabla_uu$, FW transport reads $\nabla_uV=-\epsilon[(a\cdot V)u-(u\cdot V)a]$. On $\gamma_r$ the tetrad components $T^a=(H,0,D,0)$ are constant and $a^a=(\Gamma_\phi u)^a/|T|$. Nolan~\cite{Nolan} solved this problem for general stationary, cylindrically symmetric metrics; his master frequency, Eq.~(3.12) of Ref.~\cite{Nolan}, applies directly here. We rederive it in the $so(2,1)$ language of Sec.~\ref{sec:circles}, which makes the comparison with parallel transport immediate. The FW generator is a constant element of $so(2,1)$,
\begin{equation}\label{GFW}
G_{\rm FW}=-\Gamma_\phi-\frac{1}{T\cdot T}\Big[T\,(\Gamma_\phi T)^{\!\top}-(\Gamma_\phi T)\,T^{\top}\Big]\eta ,
\end{equation}
with $G_{\rm FW}T=0$. Multiplying out the matrices, with $\eta=\operatorname{diag}(1,-1,-1)$ on the $(0,1,2)$ block and $T\cdot T=H^2-D^2$, we obtain
\begin{equation}
\label{theta0}
\theta^2\equiv-\frac12\operatorname{tr}G_{\rm FW}^2
=\frac{(AD+BH)^2}{H^2-D^2}.
\end{equation}
On a timelike circular orbit $H^2-D^2>0$, and $\theta$ fixes the angle $2\pi\theta$ by which the gyroscope rotates, in the observer's rest space, on each loop. Substituting the functions $A$, $B$, $H$, and $D$ gives
\begin{equation}\label{theta}
\theta^2=\frac{k\,(2k-1)\,(C-1)^3}{2\,(C-C_c)} ,
\end{equation}
This is positive exactly on the CTCs ($C>C_c$) and blows up as the curve becomes null. For $r<r_c$ the circle is spacelike and Eq.~(\ref{theta}) is negative; the FW map is then a boost in the Lorentzian normal space. Evaluating Nolan's general expression for this family gives the same result, as it must. For SR and G\"odel it reduces to
\begin{align}
\text{SR:}&\quad \theta=\frac{x^3}{\sqrt{x^2-1}},\qquad \theta_{\min}=\frac{3\sqrt3}{2}\ \text{at}\ x=\sqrt{3/2},\label{thetaSR}\\
\text{G\"odel:}&\quad \theta^2=\frac{(C-1)^3}{2(C-3)},\qquad \theta_{\min}=3\sqrt{3/2}\ \text{at}\ C=4.\label{thetaG}
\end{align}
These coincide with Eqs.~(3.19) and (3.23) of Ref.~\cite{Nolan} (the homogeneous SR metric is the case $b=1$ of his charged-dust family, and his G\"odel scale is $\alpha=m/2$), including the minima $\theta^2_{\min}=27/4$ and $27/2$. A gyroscope taken once around a circular CTC therefore never comes back unrotated, a spin along $e^3$ is always periodic, and the other two directions return only if $\theta\in\mathbb{Z}$, which happens on two CTCs for each integer $\theta\ge3$ in SR and $\theta\ge4$ in G\"odel (Fig.~\ref{fig:FW}). What Eq.~(\ref{theta}) adds is the dependence on $k$ across the whole family: $\theta^2$ is positive exactly on the CTCs, so for $k\le1/2$, where there are no CTCs, the question does not arise, and for every $k>1/2$ the same structure appears, with a minimum that depends on $k$.

The contrast with Sec.~\ref{sec:circles}, which is not discussed in Ref.~\cite{Nolan}, is worth noting. Just outside the chronology horizon, parallel transport around a CTC is a boost (Proposition~\ref{prop2}), yet the gyroscope simply rotates. In SR this holds throughout the band $1<x<\varphi$. The difference comes from the acceleration term in Eq.~(\ref{GFW}), which supplies the boost needed to keep $u$ fixed.

\section{Superfluid analog}\label{sec:analogue}

This section depends on an assumption that should be stated up front. A fluid in rigid rotation, treated with the standard acoustic metric~\cite{BLV}, is just flat space in rotating coordinates; it has no vorticity in the geometric sense and no CTCs. The SR form used below comes instead from the coarse-grained description of a vortex lattice proposed by Chapline and Mazur~\cite{ChaplineMazur}, and our conclusions hold to the extent that this description holds.

According to Chapline and Mazur~\cite{ChaplineMazur}, the region outside the core of a single superfluid vortex is described by the spinning-cosmic-string metric $ds^2=(c_sdt+\xi d\phi)^2-dr^2-r^2d\phi^2-dz^2$, with $\xi=\kappa/2\pi c_s$ and $\kappa=h/M$ the circulation quantum. A dense lattice of vortices with areal density $\sigma$ rotates rigidly at $\Omega=\kappa\sigma/2$ (Feynman's relation~\cite{Feynman}), and on large scales it is described by the SR form
\begin{equation}\label{acoustic}
ds^2=\Big(c_sdt+\frac{\Omega}{c_s}r^2d\phi\Big)^2-dr^2-r^2d\phi^2-dz^2 .
\end{equation}
Everything in Secs.~\ref{sec:circles}--\ref{sec:FW} carries over with $x=\Omega r/c_s$. The effective CTC radius $r_c=c_s/\Omega$ is simply the radius at which the rigid rotation becomes supersonic. Chapline and Mazur link the appearance of CTCs there to the breakdown of superfluid rigidity~\cite{ChaplineMazur}.

\emph{Orders of magnitude.} In superfluid $^4$He ($c_s\simeq238$~m/s) rotating at $\Omega\sim1$~s$^{-1}$, $r_c$ is of order 100~m, far larger than any container. Dilute Bose--Einstein condensates are more promising, but there is a real tension. The coarse-grained metric (\ref{acoustic}) makes sense only if many vortices sit inside $r_c$. With $\sigma=2\Omega/\kappa$ their number is
\begin{equation}\label{Nc}
N_c=\pi r_c^2\sigma=\frac{2\pi c_s^2}{\kappa\,\Omega},
\end{equation}
so $N_c\gg1$ requires slow rotation, $\Omega\ll2\pi c_s^2/\kappa$, while fitting the hyperbolic band inside a cloud of radius $R$ requires $\Omega\gtrsim\varphi\,c_s/R$. For $^{87}$Rb ($\kappa\simeq4.6\times10^{-9}$~m$^2$/s) with $c_s\simeq1$~mm/s, the first condition gives $\Omega\ll1.4\times10^3$~s$^{-1}$. For a uniform condensate in a box trap~\cite{Gaunt} with $R\sim50~\mu$m, somewhat larger than current traps, the second gives $\Omega\gtrsim30$~s$^{-1}$. In the window $\Omega\simeq30$--$100$~s$^{-1}$ one has $r_c\simeq10$--$30~\mu$m and $N_c\simeq15$--$45$ vortices inside the CTC radius. This is enough for a coarse-grained picture to be plausible, but not by a wide margin. The inhomogeneity of $c_s$ in a harmonic trap would also blur the critical radii, which is why a flat-bottomed trap is preferable.

\emph{What can be measured.} Phonons are scalars, so for them only the $U(1)$ part of the holonomy has physical meaning. In a metric of the form $(c_sdt+\mathcal{A})^2-h_{ij}dx^idx^j$, sound sent around a closed loop with and against the rotation arrives with a time difference $\delta t=2\oint\mathcal{A}/c_s$. For a mode of frequency $\varpi$ on a circle of radius $r$, the resulting phase difference is
\begin{equation}\label{Sagnac}
\Delta\varphi_S=\frac{4\varpi\,\Omega\,\mathcal{S}}{c_s^2},\qquad \mathcal{S}=\pi r^2,
\end{equation}
This is the acoustic Sagnac effect: the gravitational Aharonov--Bohm phase of the Kaluza--Klein-like potential $\mathcal{A}$~\cite{Sakurai,FurtadoKK}. The full $SO(2,1)$ holonomies of Sec.~\ref{sec:circles}, for vectors and spinors, matter only for quasiparticles that carry an internal frame. One example is the fermionic quasiparticles of superfluid $^3$He-A, which move as Weyl fermions in an emergent tetrad~\cite{Volovik} and acquire geometric phases when scattered by vortexlike defects~\cite{GarciaAndrade3He,Carvalho3He}; another is the polarized excitations in the analog systems reviewed in Ref.~\cite{BLV}. For these, Eqs.~(\ref{Uspin})--(\ref{W12}) predict an interference phase $\pi\sqrt{\Delta}$ that stops oscillating and starts growing exponentially at the golden-ratio radii. We stress that this is a prediction of the effective geometry; whether the corresponding excitations survive on scales of a few vortex spacings is a separate question that we do not address.

\section{Scalar field and Landau levels}\label{sec:KG}

For a scalar field of mass $\mu$ in the SR metric (\ref{SR}), $(\Box+\mu^2)\Psi=0$ with $\Psi=f(r)\,e^{-iEt+il\phi+ikz}$ and $\sqrt{-g}=r$, $g^{tt}=1-\Omega^2r^2$, $g^{t\phi}=\Omega$, $g^{\phi\phi}=-1/r^2$, gives
\begin{equation}\label{radial}
f''+\frac{f'}{r}-\frac{l^2}{r^2}f-E^2\Omega^2r^2f+\big(E^2-2E\Omega l-k^2-\mu^2\big)f=0 .
\end{equation}
This is a two-dimensional isotropic oscillator with an effective magnetic frequency $E\Omega$. Its regular, normalizable solutions are
\begin{equation}\label{wf}
f(r)=\mathcal{N}\,\rho^{|l|/2}e^{-\rho/2}\,M\big(-n_r,|l|+1,\rho\big),\qquad \rho=E\Omega r^2 ,
\end{equation}
where $M$ is Kummer's confluent hypergeometric function and $n_r=0,1,2,\ldots$. They exist only if
\begin{equation}\label{spectrum}
E=\Omega N+\sqrt{\Omega^2N^2+k^2+\mu^2},\qquad N=2n_r+|l|+l+1 .
\end{equation}
The $E<0$ branch describes antiparticles, with $l\to-l$. For $l\le0$ the energy does not depend on $l$; this is the infinite Landau degeneracy of G\"odel-type spacetimes~\cite{DFS2,FST,CarvalhoEPJC}. Equations~(\ref{wf})--(\ref{spectrum}) agree with Eqs.~(3.4)--(3.8) of Drukker, Fiol, and Sim\'on~\cite{DFS2}, up to the sign convention for $l$. For a massless field moving in the plane ($k=0$) the levels are $E=2\Omega N$.

\emph{Landau levels and the CTC radius.} Drukker, Fiol, and Sim\'on noted that the lowest-weight state of each Landau level stays inside the chronologically safe region, while the lowest-Landau-level wavefunctions, apart from the first few, peak outside it~\cite{DFS2}. That statement can be made quantitative. From Eq.~(\ref{wf}), the fraction of the radial $L^2$ norm of a lowest-Landau-level mode ($n_r=0$, $l\le0$) that lies outside this radius is
\begin{equation}\label{Pout}
P_l=\frac{\Gamma\big(|l|+1,\,E/\Omega\big)}{\Gamma\big(|l|+1\big)},
\end{equation}
since $\rho(r_c)=E/\Omega$. For a massless field with $k=0$ ($E=2\Omega$) this gives $P_0=e^{-2}\simeq0.14$, $P_{-1}=3e^{-2}\simeq0.41$, and $P_{-2}=5e^{-2}\simeq0.68$. A state peaks at $\sqrt{|l|/E\Omega}$, which is inside $r_c$ only when $|l|<E/\Omega$. For the massless case, the scale $E/\Omega=2$ means that only the first few lowest-Landau-level radial profiles are concentrated mainly inside the chronology radius; higher negative-angular-momentum modes have most of their radial $L^2$ norm outside it. This statement concerns the localization of the mode profiles; it does not count a number of independent states confined to the disk. A mass or a longitudinal momentum raises the threshold to $E/\Omega=1+\sqrt{1+(k^2+\mu^2)/\Omega^2}$. The phase $E\oint H\,d\phi=2\pi E\Omega r^2$ around a circle has the same dependence on the enclosed area as the acoustic Sagnac phase of Eq.~(\ref{Sagnac}), once the frequency and sound-speed conventions are matched. This gives a useful kinematic comparison with the Landau structure of Eq.~(\ref{spectrum}); it is not an independent derivation of the spectrum, which would require specifying a quantization rule for the phase and its assumptions.

\subsection{Exact localization beyond the lowest Landau level}
\label{subsec:loc}

\begin{figure*}[t]
\centering
\includegraphics[width=\textwidth]{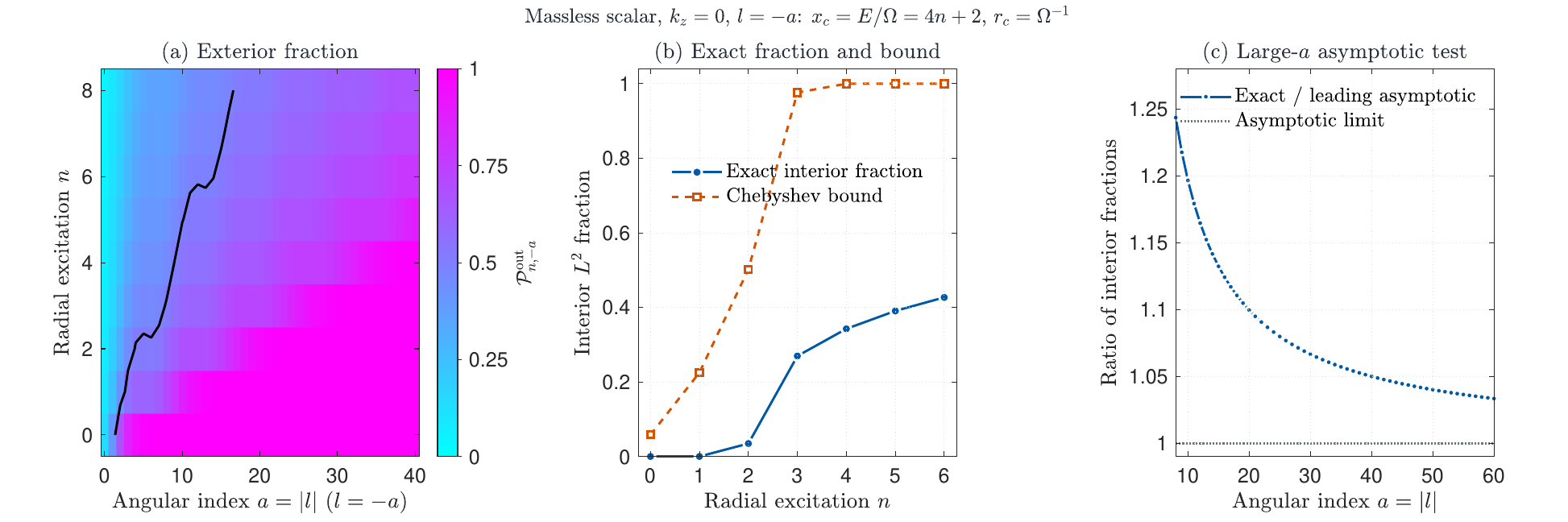}
\caption{Radial $L^2$ localization relative to the Som--Raychaudhuri chronology radius \(r_c=\Omega^{-1}\), for a massless scalar with \(k_z=0\) and \(l=-a\leq0\). The dimensionless threshold is \(x_c=E/\Omega=4n+2\). (a) Exact exterior fraction \(\mathcal P_{n,-a}^{\mathrm{out}}\), given by Eq.~(\ref{eq:locsum}), for \(n=0,\ldots,8\) and \(a=0,\ldots,40\); the contour marks \(\mathcal P_{n,-a}^{\mathrm{out}}=1/2\). (b) Exact interior fraction \(\mathcal P_{n,-a}^{\mathrm{in}}=1-\mathcal P_{n,-a}^{\mathrm{out}}\) and its Chebyshev upper bound, Eq.~(\ref{eq:locbound}), for \(a=20\) and \(n=0,\ldots,6\). (c) Ratio of the exact lowest-level interior fraction, obtained from Eq.~(\ref{Pout}), to its leading large-\(a\) asymptotic expression at fixed \(x_c\), Eq.~(\ref{eq:locasym}), for \(x_c=2\) and \(a=8,\ldots,60\); the limiting result is Eq.~(\ref{eq:loclim}).}
\label{fig:localization}
\end{figure*}

The lowest-level result in Eq.~(\ref{Pout}) has an exact extension to arbitrary radial excitation. Where a radial mode sits relative to the chronology radius is not fixed by the energy spectrum alone: it also depends on the polynomial structure of the mode and on the energy-dependent radial threshold. We therefore define an exterior localization fraction for every radial and angular quantum number, give a finite representation of it in terms of the upper incomplete gamma function, and derive a moment bound on the interior fraction. Let \(n\equiv n_r\), \(a\equiv |l|\), \(x=E_{nl}\Omega r^2\) and \(x_c=\frac{E_{nl}}{\Omega}\), where \(E_{nl}\) is given in Eq.~(\ref{spectrum}) and the chronology radius of the Som--Raychaudhuri geometry is \(r_c=\Omega^{-1}\). The radial profile in Eq.~(\ref{wf}) is proportional to \(x^{a/2}e^{-x/2}L_n^{(a)}(x)\). With respect to the radial \(L^2\) measure \(r\,dr\), its normalized distribution in \(x\) is
\begin{equation}\label{eq:locpdf}
\begin{split}
 &p_{na}(x)=\frac{n!}{\Gamma(n+a+1)}
x^a e^{-x}\big[L_n^{(a)}(x)\big]^2,\\
&\int_0^\infty p_{na}(x)\,dx=1.   
\end{split}
\end{equation}
The normalization follows from the orthogonality of the associated Laguerre polynomials with weight \(x^a e^{-x}\)~\cite[Sec.~18.3]{DLMF}. The exterior fraction is then
\begin{equation}
\begin{split}
\mathcal P_{nl}^{\mathrm{out}}
={}&\frac{n!}{\Gamma(n+a+1)}
\sum_{i=0}^{n}\sum_{j=0}^{n}
\frac{(-1)^{i+j}}{i!\,j!}
\binom{n+a}{n-i}
\\
&\quad\times
\binom{n+a}{n-j}
\Gamma(a+i+j+1,x_c).\label{eq:locsum}
\end{split}
\end{equation}
Figure~\ref{fig:localization}(a) shows this exterior fraction for \(n=0,\ldots,8\) and \(a=0,\ldots,40\); the contour marks \(\mathcal P_{n,-a}^{\mathrm{out}}=1/2\).
The variance of this distribution is
\begin{equation}
\label{eq:locvar}
\operatorname{Var}_{na}(x)
=(n+1)(n+a+1)+n(n+a).
\end{equation}
The mean of the distribution is \(2n+a+1\). For \(x_c<2n+a+1\), Chebyshev's inequality~\cite{Billingsley} then gives
\begin{equation}
\begin{split}
\mathcal P_{nl}^{\mathrm{in}}&\leq \min\!\left\{1,\,\frac{(n+1)(n+a+1)+n(n+a)}
{(2n+a+1-x_c)^2}\right\},\label{eq:locbound}
\end{split}
\end{equation}
where \(\mathcal P_{nl}^{\mathrm{in}}=1-\mathcal P_{nl}^{\mathrm{out}}\). This bound is rigorous but loses sharpness as the threshold approaches the mean; it is most useful when the mean lies well beyond the threshold. Figure~\ref{fig:localization}(b) compares the exact interior fraction with the Chebyshev bound for \(a=20\) and \(n=0,\ldots,6\). The lowest Landau level with \(l=-a\leq0\) gives a particularly transparent limit. In this sector the energy, and hence \(x_c\), does not depend on \(a\), and the incomplete-gamma expression in Eq.~(\ref{Pout}) implies
\begin{equation}
\lim_{a\to\infty}\mathcal P_{0,-a}^{\mathrm{out}}=1.\label{eq:loclim}
\end{equation}
For fixed \(x_c>0\), the complementary interior fraction has the large-\(a\) expansion
\begin{equation}
\begin{split}
\mathcal P_{0,-a}^{\mathrm{in}}=\frac{e^{-x_c}x_c^{a+1}}{\Gamma(a+2)}
\left[1+O\!\left(\frac{x_c}{a+2}\right)\right],\quad a\to\infty. \label{eq:locasym}
\end{split}
\end{equation}
This follows from the series expansion of the normalized lower incomplete gamma function~\cite[Eq.~(8.7.1)]{DLMF}. The estimate assumes that \(x_c\) stays fixed as \(a\) grows; it is not uniform when the threshold grows with the quantum number. It quantifies how the lowest-level radial \(L^2\) distribution concentrates outside the chronology radius at large negative angular momentum. Figure~\ref{fig:localization}(c) compares the leading asymptotic expression at fixed \(x_c\) with the exact interior fraction for \(x_c=2\) and \(a=8,\ldots,60\). These results do not, by themselves, establish a dynamical instability, chronology protection, or a probability interpretation of a Klein--Gordon mode. Equations~(\ref{eq:locsum})--(\ref{eq:locbound}) thus provide two complementary results, an exact finite representation for arbitrary radial excitation and a rigorous moment bound on the interior fraction. Together with the lowest-level limit (\ref{eq:loclim}), they extend the comparison between the Landau spectrum and the chronology radius.

All panels of Fig.~\ref{fig:localization} refer to a massless scalar with \(k_z=0\) and \(l=-a\leq0\). In this case \(E=2\Omega(2n+1)\), so \(x_c=E/\Omega=4n+2\), while the Som--Raychaudhuri chronology radius is \(r_c=\Omega^{-1}\).

\section{Concluding remarks}\label{sec:conclusions}

We have classified the parallel-transport holonomies of circles centered on the symmetry axis in the homogeneous G\"odel-type family in terms of a single Lorentz-invariant quantity. Because every connection component lies in $so(2,1)$, the holonomy depends on a single invariant, $\Delta(r)=B^2-A^2$. Its sign tells whether an orbit is elliptic, parabolic, or hyperbolic, and its closed form (\ref{Delta}) divides the family into three classes, $k\le1/2$, $1/2<k\le2$, and $k>2$, with zero, one, and two parabolic radii. The result we find most interesting is Proposition~\ref{prop2}. Whenever CTCs exist, the circle where they first appear has $\Delta=-\cosh mr_c$, so it always lies in a hyperbolic band. Equation~(\ref{DeltaKilling}) explains why: $\Delta$ depends only on the two Killing scalars $u\cdot\xi$ and $\xi\cdot\xi$, and on the horizon, where $\xi\cdot\xi=0$, the synchronization gap $u\cdot\xi$ is always large enough to turn the holonomy into a boost. In the Som--Raychaudhuri limit the edges of that band are the golden-ratio radii $\Omega r=\varphi^{\mp1}$.

Along the way we corrected the condition for band holonomy invariance, Eqs.~(\ref{band})--(\ref{bandSR}), separated the vector and spinor cases, and showed that hyperbolic bands contain no orbits with trivial holonomy. We also showed that the precession seen along time displacements comes from the vorticity, not from the CTCs. On the CTCs themselves, the Fermi--Walker analysis of Nolan~\cite{Nolan} supplies the measurable counterpart. We extended it to the whole family and showed how the boost-type parallel holonomy near the chronology horizon is converted into a pure rotation of the gyroscope by the observer's acceleration.

In the superfluid analog the $U(1)$ part of the holonomy is the acoustic Sagnac phase, and the critical radii scale as $c_s/\Omega$. For uniform condensates there is a narrow window, $r_c\sim10$--$30~\mu$m with a few tens of vortices inside, where the effective description is plausible but not guaranteed. On the quantum side, we made quantitative the observation of Ref.~\cite{DFS2}: for a massless field, only the first few lowest-Landau-level profiles are concentrated mainly inside the chronology radius. The localization analysis extends this comparison to arbitrary radial excitation, through an exact finite representation of the exterior fraction of the norm and a rigorous moment bound on the interior fraction. In the lowest Landau level the exterior fraction tends to unity at large negative angular momentum for a fixed radial threshold. These statements concern the radial $L^2$ distribution; by themselves they establish neither a probability interpretation of a Klein--Gordon mode nor a dynamical instability or chronology protection. Several extensions are within reach. Adding topological defects such as cosmic strings or dislocations to the G\"odel-type background~\cite{CarvalhoEPJC,GarciaTorsion,GarciaWeyl,BouzenadaSR} would multiply the holonomy by a conical factor. Circles not centered on the axis, including the closed geodesics, remain to be studied. The Dirac equation should show the spinor holonomy (\ref{Uspin}) as spin-dependent Landau levels.

\begin{acknowledgments}
This work was supported by Conselho Nacional de Desenvolvimento Cient\'{\i}fico e Tecnol\'ogico (CNPq) and Funda\c{c}\~ao de Apoio \`a Pesquisa do Estado da Para\'iba (Fapesq-PB). G. Q. Garcia acknowledges financial support from Fapesq-PB (Grant BLD-ADT-A2377/2024). C. Furtado acknowledges support from CNPq (PQ Grant 1A No. 311781/2021-7).

\section*{Declaration of AI use}
The authors used an AI assistant for language editing and for independent symbolic checks of the calculations. The authors defined the research problem, selected the results to be explored and included, and provided and reviewed the relevant prior literature. They verified all derivations by hand and checked them with computer algebra, as described in the Supplemental Material, and they take full responsibility for the scientific content of this work.
\end{acknowledgments}

\section*{Data availability}
No experimental data were generated in this theoretical study. The computer-algebra scripts used to check the results, and the scripts that generate the figures, are provided in the Supplemental Material~\cite{SM}.


\begin{thebibliography}{99}

\bibitem{Mandelstam} S. Mandelstam, Ann. Phys. (N.Y.) \textbf{19}, 25 (1962).

\bibitem{Mandelstam1} S. Mandelstam, Phys. Rev. \textbf{175}, 1580 (1968).

\bibitem{Yang} C. N. Yang, Phys. Rev. Lett. \textbf{33}, 445 (1974).

\bibitem{Voronov} N. A. Voronov and Yu. M. Makeenko, Sov. J. Nucl. Phys. \textbf{36}, 444 (1982).

\bibitem{Wilson} K. G. Wilson, Phys. Rev. D \textbf{10}, 2445 (1974).

\bibitem{Ashtekar} A. Ashtekar, Phys. Rev. Lett. \textbf{57}, 2244 (1986).

\bibitem{RovelliSmolin1} C. Rovelli and L. Smolin, Phys. Rev. Lett. \textbf{61}, 1155 (1988).

\bibitem{RovelliSmolin2} C. Rovelli and L. Smolin, Nucl. Phys. B \textbf{331}, 80 (1990).

\bibitem{BGT} C. G. Bollini, J. J. Giambiagi and J. Tiomno, Lett. Nuovo Cimento \textbf{31}, 13 (1981).

\bibitem{Rothman} T. Rothman, G. F. R. Ellis and J. Murugan, Class. Quantum Grav. \textbf{18}, 1217 (2001).

\bibitem{Bini} D. Bini, C. Cherubini and R. T. Jantzen, Class. Quantum Grav. \textbf{19}, 5481 (2002).

\bibitem{CarvalhoBS} A. M. de M. Carvalho, F. Moraes and C. Furtado, Class. Quantum Grav. \textbf{20}, 2063 (2003).

\bibitem{CarvalhoCigar} A. M. de M. Carvalho, C. Furtado and F. Moraes, J. High Energy Phys. \textbf{06} (2004) 029.

\bibitem{CarvalhoFRW} A. M. de M. Carvalho and C. Furtado, Gen. Relativ. Gravit. \textbf{39}, 1311 (2007).

\bibitem{Burgers} C. J. C. Burges, Phys. Rev. D \textbf{32}, 504 (1985).

\bibitem{Bezerra} V. B. Bezerra, Phys. Rev. D \textbf{35}, 2031 (1987).

\bibitem{FordVilenkin} L. H. Ford and A. Vilenkin, J. Phys. A \textbf{14}, 2353 (1981).

\bibitem{Assis1} J. G. de Assis, C. Furtado, V. B. Bezerra and F. Moraes, Grav. Cosmol. \textbf{6}, 233 (2000).

\bibitem{Assis2} J. G. de Assis, C. Furtado and V. B. Bezerra, Phys. Rev. D \textbf{62}, 045003 (2000).

\bibitem{BezerraLetelier} V. B. Bezerra and P. S. Letelier, J. Math. Phys. \textbf{37}, 6271 (1996).

\bibitem{FurtadoKK} C. Furtado, F. Moraes and V. B. Bezerra, Phys. Rev. D \textbf{59}, 107504 (1999).

\bibitem{Godel} K. G\"odel, Rev. Mod. Phys. \textbf{21}, 447 (1949).

\bibitem{Hawking} S. W. Hawking, Phys. Rev. D \textbf{46}, 603 (1992).

\bibitem{RT} M. J. Rebou\c{c}as and J. Tiomno, Phys. Rev. D \textbf{28}, 1251 (1983).

\bibitem{SomRay} M. M. Som and A. K. Raychaudhuri, Proc. R. Soc. A \textbf{304}, 81 (1968).

\bibitem{HorowitzTseytlin} G. T. Horowitz and A. A. Tseytlin, Phys. Rev. D \textbf{51}, 2896 (1995).

\bibitem{RussoTseytlin1} J. G. Russo and A. A. Tseytlin, Nucl. Phys. B \textbf{448}, 293 (1995).

\bibitem{RussoTseytlin2} J. G. Russo and A. A. Tseytlin, Nucl. Phys. B \textbf{454}, 164 (1995).

\bibitem{Boyda} E. K. Boyda, S. Ganguli, P. Ho\v{r}ava and U. Varadarajan, Phys. Rev. D \textbf{67}, 106003 (2003).

\bibitem{Harmark} T. Harmark and T. Takayanagi, Nucl. Phys. B \textbf{662}, 3 (2003).

\bibitem{DFS1} N. Drukker, B. Fiol and J. Sim\'on, Phys. Rev. Lett. \textbf{91}, 231601 (2003).

\bibitem{DFS2} N. Drukker, B. Fiol and J. Sim\'on, J. Cosmol. Astropart. Phys. \textbf{10} (2004) 012.

\bibitem{DasGegenberg} S. Das and J. Gegenberg, Gen. Relativ. Gravit. \textbf{40}, 2115 (2008).

\bibitem{FST} B. D. B. Figueiredo, I. Dami\~ao Soares and J. Tiomno, Class. Quantum Grav. \textbf{9}, 1593 (1992).

\bibitem{CarvalhoEPJC} J. Carvalho, A. M. de M. Carvalho and C. Furtado, Eur. Phys. J. C \textbf{74}, 2935 (2014).

\bibitem{GarciaTorsion} G. Q. Garcia, J. R. de S. Oliveira, K. Bakke and C. Furtado, Eur. Phys. J. Plus \textbf{132}, 123 (2017).

\bibitem{GarciaWeyl} G. Q. Garcia, J. R. de S. Oliveira and C. Furtado, Int. J. Mod. Phys. D \textbf{27}, 1850027 (2018).

\bibitem{VitoriaSR} R. L. L. Vit\'oria, C. Furtado and K. Bakke, Eur. Phys. J. C \textbf{78}, 44 (2018).

\bibitem{BouzenadaSR} A. Bouzenada, A. Boumali, R. L. L. Vit\'oria and C. Furtado, Theor. Math. Phys. \textbf{221}, 2193 (2024).

\bibitem{ChaplineMazur} G. Chapline and P. O. Mazur, Acta Phys. Pol. B \textbf{45}, 905 (2014); arXiv:gr-qc/0407033.

\bibitem{Edinelson} E. P. dos Santos, \emph{Vari\'aveis de contorno e loop de Wilson para uma classe de espa\c{c}os-tempo}, M.Sc. dissertation, Universidade Federal da Para\'iba (2011).

\bibitem{BiniGodel} D. Bini, A. Geralico, R. T. Jantzen and W. Plastino, Phys. Rev. D \textbf{100}, 084051 (2019).

\bibitem{Nolan} B. C. Nolan, Phys. Rev. D \textbf{104}, 064041 (2021).

\bibitem{SM} See Supplemental Material at [URL will be inserted by publisher] for the verification scripts and a description of the checks performed.

\bibitem{BakkeFW} K. Bakke, A. M. de M. Carvalho and C. Furtado, Eur. Phys. J. C \textbf{63}, 149 (2009).

\bibitem{BLV} C. Barcel\'o, S. Liberati and M. Visser, Living Rev. Relativ. \textbf{14}, 3 (2011).

\bibitem{Feynman} R. P. Feynman, in \emph{Progress in Low Temperature Physics}, edited by C. J. Gorter (North-Holland, Amsterdam, 1955), Vol.~1, p.~17.

\bibitem{Gaunt} A. L. Gaunt, T. F. Schmidutz, I. Gotlibovych, R. P. Smith and Z. Hadzibabic, Phys. Rev. Lett. \textbf{110}, 200406 (2013).

\bibitem{Sakurai} J. J. Sakurai, Phys. Rev. D \textbf{21}, 2993 (1980).

\bibitem{Volovik} G. E. Volovik, \emph{The Universe in a Helium Droplet} (Oxford University Press, Oxford, 2003).

\bibitem{GarciaAndrade3He} L. C. Garcia de Andrade, A. M. de M. Carvalho and C. Furtado, Europhys. Lett. \textbf{67}, 538 (2004).

\bibitem{Carvalho3He} A. M. de M. Carvalho, A. S. Marques, G. T. Silva, G. Q. Garcia and C. Furtado, Int. J. Geom. Methods Mod. Phys. \textbf{22}, 2550115 (2025).

\bibitem{DLMF} F. W. J. Olver, A. B. Olde Daalhuis, D. W. Lozier, B. I. Schneider, R. F. Boisvert, C. W. Clark, B. R. Miller, and B. V. Saunders, eds., \emph{NIST Digital Library of Mathematical Functions}, https://dlmf.nist.gov/ (National Institute of Standards and Technology, Gaithersburg, MD).

\bibitem{Billingsley}
P. Billingsley, \emph{Probability and Measure}, 3rd ed.
(Wiley, New York, 1995).
\end{thebibliography}
\end{document}